\documentclass[twocolumn,showpacs,pra,aps]{revtex4-1}

\usepackage[T1]{fontenc}     
\usepackage[british]{babel}  
\usepackage{lmodern}  

\usepackage[usenames,dvipsnames]{color} 

\usepackage{graphicx} 
\usepackage{amsmath,amssymb,amsthm,bm,amsfonts,mathrsfs,bbm} 
\usepackage{xspace}  
\usepackage{pgfplots}  
\usepackage[colorlinks,linkcolor=magenta,urlcolor=blue]{hyperref}  
\usepackage{mathtools}

\DeclareMathOperator{\Tr}{Tr}

\newcommand{\ba}{\begin{eqnarray}}
\newcommand{\ea}{\end{eqnarray}}
\newcommand{\ban}{\begin{eqnarray*}}
\newcommand{\ean}{\end{eqnarray*}}

\def\one{\leavevmode\hbox{\small1\normalsize\kern-.33em1}}

\begin{document}


\title{Family of Bell inequalities violated by higher-dimensional bound entangled states}

\author{K\'aroly F. P\'al}
\affiliation{Institute for Nuclear Research, Hungarian Academy of Sciences, H-4001 Debrecen, P.O. Box 51, Hungary}

\author{Tam\'as V\'ertesi}
\affiliation{Institute for Nuclear Research, Hungarian Academy of Sciences, H-4001 Debrecen, P.O. Box 51, Hungary}

\date{\today}  

\begin{abstract}
We construct ($d\times d$)-dimensional bound entangled states, which
violate, for any $d>2$, a bipartite Bell inequality introduced in
this paper. We conjecture that the proposed class of Bell
inequalities acts as a dimension witness for bound entangled
states: For any $d>2$ there exists a Bell inequality from this
class that can be violated with bound entangled states only if
their Hilbert space dimension is at least $d\times d$. Numerics
supports this conjecture up to $d=8$.
\end{abstract}

\maketitle

\section{Introduction}
\label{Intro}

Distant parties carrying out suitable local measurements on a shared quantum state can establish nonlocal correlations, which are signaled by the violation of Bell inequalities~\cite{bell64,bellreview1,bellreview2}. A Bell violation implies that the underlying quantum state is entangled. Such a violation of a Bell inequality has been attained in recent experiments simultaneously closing both main technical loopholes, the so-called locality and the detection loopholes~\cite{EXP1,EXP2,EXP3,EXP4}.

However, from the theoretical point of view, it is still unknown whether all entangled quantum states can lead to violation of a Bell inequality~\cite{bellreview1,bellreview2,horoRMP}. For instance, there exist mixed two-qubit entangled states, so-called Werner states~\cite{werner}, which admit a local hidden-variable model for any general one-shot measurement and hence cannot violate any Bell inequality~\cite{barrett} (see also more recent related results in Refs.~\cite{betterhirsch,simpovm}).

However, in the case of more general scenarios, nonlocality of certain mixed states can be activated. Such alternative scenarios involve the multicopy case, i.e., when multiple copies of a given quantum state can be measured jointly in a Bell test~\cite{chshactivation,palazuelos,telep}, and the case where preprocessing using local operations and classical communications (LOCC) can be carried out on the state before performing the Bell test itself~\cite{popescu,masanes}.

In the most general case, both above actions are allowed prior to a Bell text, that is, any number of copies of the state in question can be preprocessed by means of LOCC operations. In this way, the problem of nonlocality of quantum states becomes closely related to the task of entanglement distillation~\cite{BE,ED}. In such a protocol, one starts from an arbitrary number of copies of a state and tries to extract a pure highly entangled state by LOCC. Put together, it follows that any entangled state that is distillable can give rise to Bell inequality violation.

Indeed, it has been shown in 1998 that there exist entangled states in nature that are not distillable~\cite{BE}. Such a prominent class of entangled states is the so-called bound entangled states. From this type of state, it is not possible to distill pure maximally entangled states by LOCC. On the other hand, entangled states are required to produce these states. This kind of irreversible behavior of bound entangled states represents a very weak form of entanglement, which led Peres to conjecture in 1999 that bound entanglement can never lead to Bell inequality violation~\cite{peres}.

Several results have been reported in favor of the Peres conjecture~\cite{WW,Toni,Lluis,Dani,Pusey,TobiasDI,LHVconst}. However, it has been refuted recently in Ref.~\cite{VB}, where the presented nonlocal bound entangled state is a member of the family of $3\times 3$ bound entangled states from Ref.~\cite{tobias}. The bound entangled states of this family are positive with respect to the partial transpose~\cite{PPT}, and a subset of these states has been shown~\cite{VB} to violate the Pironio-Bell inequality~\cite{pironio}. It is noted that the multipartite version of the Peres conjecture, which is a weaker version of his original conjecture, has also been addressed~\cite{dur,augusiak,VB12,ruben}.

In this paper we address the question of the existence of nonlocal higher-dimensional bound entangled states. In particular, we propose ($d\times d$)-dimensional positive partial transpose (PPT) bound entangled states for any $d>2$ that violate a class of bipartite Bell inequalities. The setup involves $d$ binary-outcome measurements on Alice's side and one $d$-outcome and one binary-outcome measurement on Bob's side. The constructed states are invariant with respect to partial transposition, a property that ensures that they are PPT states. The rank of these states is $2d-2$. Note that Ref.~\cite{pptrank} conjectures that this is the lowest possible rank among $d\times d$ extremal partial transpose invariant states. On the other hand, we conjecture that our inequalities act as dimension witnesses for bound entangled states: For any $d>2$ there exists a member from our class of Bell inequalities that cannot be violated with [$(d-1)\times (d-1)$] PPT entangled states. However, they can be violated with our $d\times d$ PPT entangled quantum systems. The constructions presented in the paper can be considered as a straightforward generalization of the PPT state (and the Pironio-Bell inequality) in Ref.~\cite{VB} for any $d>3$. Provided our conjecture is true, it can also be considered as a device-independent dimension witness~\cite{dimwit} for bound entangled states of any dimension $d>2$.

\section{Previous work}
\label{previous}

In a recent paper Yu and Oh~\cite{YO} have given a family of nonlocal bipartite bound entangled states. Each member of the family is defined by a density matrix in a $(d\times d)$-dimensional ($d\geq 3$) Hilbert space. The states are invariant under partial transposition; consequently, they are PPT states~\cite{PPT,horoRMP}. The rank of the state characterized by $d$ is $d(d-1)/2+1$. Yu and Oh have proven the nonlocality of their states by showing that each of them can violate a Bell inequality. In the Bell scenario they have given Alice has $d$ two-outcome measurements, while Bob has one $d$-outcome and one two-outcome measurement. The Bell inequality may be written as
\begin{equation}
p(00|01)-p(00|00)-\sum_{i=1}^{d-1}p(0i|i0)-\sum_{i=1}^{d-1}p(10|i1)\leq 0,
\label{eq:yuohineq}
\end{equation}
where $p(ab|xy)$ denotes the conditional probability of Alice and Bob getting outcome $a$ and $b$, provided they have chosen settings $x$ and $y$, respectively. Both the settings and the outcomes are labeled with non-negative integers starting from zero. Note that the family of inequalities above is equivalent to the one defined by Pironio in Ref.~\cite{pironio}.

Yu and Oh~\cite{YO} have shown that if the measurement settings and the parameters of the states are chosen appropriately, the inequality given in Eq.~(\ref{eq:yuohineq}) is indeed violated. The violation decreases with $d$ and for $d$ large it is proportional to $d^{-4}$. However, it is easy to show that for $d>3$ the $d$th member of the family of states is not the maximally violating PPT states for the $d$th inequality. Let the last measurement of Alice (the one labeled $d-1$) be a degenerate one such that its outcome is always zero. Then $p(1,0|d-1,1)=0$. Also, let the first measurement of Bob be such that its last outcome never happens. Then $p(0,d-1|d-1,0)=0$. With these choices of measurements, Eq.~(\ref{eq:yuohineq}) corresponding to $d$ will be reduced to the one corresponding to $d-1$. Any state defined in the [$(d-1)\times(d-1)$]-dimensional subspace that violates the inequality of $d-1$ will equally violate the inequality of $d$. Therefore, all inequalities will be violated by at least as much as the $d=3$ one by a PPT state. For this family of inequalities the violation cannot give any information about the dimensionality of the state. In the present paper we introduce an alternative family of inequalities and a family of Bell nonlocal PPT states.

\section{Tight Bell inequalities for $d=4$}
\label{tight4ineqs}

Using the polytope software PORTA~\cite{porta}, we have generated all tight inequalities with Alice having $d=4$ two-outcome measurements and Bob having one $(d=4)$-outcome and one two-outcome measurement. We have 11136 of them. Most of these are equivalent to trivial inequalities [$-p(00|00)\leq 0$ and $-p(00|01)\leq 0$] or to the Clauser-Horne-Shimony-Holt Bell inequality~\cite{chsh}. Also, several of them are equivalent to the $d=3$ or the $d=4$ inequalities given in Eq.~(\ref{eq:yuohineq}). The rest is equivalent to one of two inequalities. To the best of our knowledge, this scenario has not been resolved so far. In particular, it is not present in the database~\cite{faacets}.

The first one of these two inequalities may be written as
\begin{align}
&p(00|01)-p(00|00)-p(02|10)-p(03|10)\nonumber\\
&-p(10|11)-p(01|20)-p(03|20)-p(10|21)\nonumber\\
&+p(03|30)-p(00|31)\leq 0.
\label{eq:alt2ineq4}
\end{align}
We have used a seesaw-type algorithm~\cite{seesaw1,seesaw2,seesaw3} similar to the one used in Ref.~\cite{VB}, which will be explained in the next section, to find the PPT state and the measurement settings violating the above inequality the most.
The maximum violation we have found this way was the same as the maximum one can get for the inequality in Eq.~(\ref{eq:yuohineq}) for $d=3$. The situation is very similar to the cases of
Eq.~(\ref{eq:yuohineq}) for $d>3$. If we take the measurement settings such that neither outcome zero of measurement three of Alice nor outcome three of measurement zero of Bob ever happens, then
$p(03|10)$, $p(03|20)$, $p(03|30)$, and $p(00|31)$ are all zero and what remains is equivalent to Eq.~(\ref{eq:yuohineq}) for $d=3$. Therefore, the same violation with the same state can always be achieved.

The other inequality may be written as
\begin{align}
2&[p(00|01)-p(00|00)]-p(02|10)-p(03|10)\nonumber\\
&-p(10|11)-p(01|20)-p(03|20)-p(10|21)\nonumber\\
&-p(01|30)-p(02|30)-p(10|31)\leq 0.
\label{eq:alt1ineq4}
\end{align}
This inequality can not be reduced to the $d=3$ inequality by choosing measurement settings such that the probabilities of some of the outcomes are zero. There exist ($4\times4$)-dimensional PPT states violating this inequality, but we could not find any PPT state in a smaller space doing that.

\section{Generalization of Bell inequalities beyond $d=4$}
\label{generald4}

\subsection{The inequality}
\label{genineq}

The last and most interesting inequality of Eq.~(\ref{eq:alt1ineq4}) may be generalized
to any $d\geq 3$ as
\begin{align}
I_d = &(d-2)[p(00|01)-p(00|00)]\nonumber\\
&-\sum_{i,j=1}^{d-1}p(0j|i0)(1-\delta_{ij})-\sum_{i=1}^{d-1}p(10|i1)\leq 0.
\label{eq:alt1ineq}
\end{align}
For $d=3$ the inequality is the same as the one of Eq.~(\ref{eq:yuohineq}), only
Alice's measurements one and two are swapped. It is not difficult to show that the
classical bound appearing on the right-hand side of the equation is indeed zero.
This number is the maximum value the left-hand side can take if the conditional probabilities
are given by deterministic strategies. In a deterministic strategy the outcome of
each measurement is certain for both parties, independently of each other.
Therefore, $p(ab|xy)=\alpha_{a|x}\beta_{b|y}$, where $\alpha_{a|x}$ ($\beta_{b|y}$), the probability of Alice (Bob) getting outcome $a$ ($b$) provided she (he) performs measurement $x$ ($y$), is one for each $x$ ($y$) for one of the outcomes, and zero for all other outcomes.

The classical bound zero can be achieved with many deterministic strategies; for example, with the choice of $\alpha_{1|i}=\beta_{0|0}=\beta_{1|1}=1$ each term on the right-hand side of Eq.~(\ref{eq:alt1ineq}) is zero.
Now we will show that we can not get a positive classical value. The only term in the equation that can give a positive contribution is the first one. For that the choice of $\alpha_{0|0}=1$ and $\beta_{0|1}=1$ has to be made.
Then this term will have the value of $d-2$. For Bob's $d$-outcome measurement zero $\beta_{0|0}=1$ would
lead to $p(00|00)=1$, which would give a contribution of  $-(d-2)$, negating the positive term. Therefore,
let us choose $\beta_{b_0|0}=1$ ($0<b_0\leq d$). Then, no matter how we choose $\alpha_{0|i}$ for $i\neq 0$, either $p(0b_0|i0)=\alpha_{0|i}\beta_{b_0|0}=1$ or $p(10|i1)=\alpha_{1|i}\beta_{0|1}=1$. This way for each $i$ we get a contribution of -1 except for $i=b_0$ provided $\alpha_{0|b_0}=1$, as $p(0b_0|b_00)$ has a zero factor. Therefore, we get at least $d-2$ terms of value $-1$, so we can not get a sum larger than zero, indeed.


\subsection{Optimization of the PPT quantum value}
\label{optPPT}

The quantum value of a conditional probability appearing in a Bell inequality may be written as
\begin{equation}
p(ab|xy)=\Tr[\hat\rho(\hat A_{a|x}\otimes\hat B_{b|y})],
\label{eq:condprob}
\end{equation}
where $\hat A_{a|x}$ and $\hat B_{b|y}$ are the operators
corresponding to outcome $a$ and $b$ of Alice's and Bob's
measurement setting $x$ and $y$, respectively. We allow
positive-operator-valued-measure (POVM) measurements. Therefore,
the quantum value of the Bell expression can be written as
\begin{equation}
Q(d)=\Tr(\hat\rho\hat{\cal B}_d),
\label{eq:qtrrB}
\end{equation}
where the Bell-operator $\hat{\cal B}$ is the linear combination of
the operators $\hat A_{a|x}\otimes\hat B_{b|y}$ according to the Bell coefficients. For
inequality $I_d$ in Eq.~(\ref{eq:alt1ineq}) it takes the form:
\begin{align}
\hat{\cal B}_d=&(d-2)\hat A_{0|0}\otimes(\hat B_{0|1}-\hat B_{0|0})\nonumber\\
&-\sum_{i,j=1}^{d-1}\hat A_{0|i}\otimes\hat B_{j|0}(1-\delta_{ij})-\sum_{i=1}^{d-1}\hat A_{1|i}\otimes\hat B_{0|1}.
\label{bellopa}
\end{align}
When the quantum value is larger than the classical bound, the inequality is violated.

From Eq.~(\ref{eq:qtrrB}) it follows that given $\hat{\cal B}$ (that is, given the measurement settings), finding the optimal density matrix is a problem of semidefinite programming (SDP)~\cite{sdp}. Confining ourselves
to PPT states is just a matter of another standard constraint in the SDP method.
It is also true that given a state and the measurement settings of one of the parties, finding the optimum settings for the other party is also an SDP problem. This is because for each measurement setting the operators corresponding to the outcomes are positive-semidefinite operators whose sum is the identity operator, and the quantum value to be maximized is a linear combination of the matrix elements of these operators.
The seesaw algorithm~\cite{seesaw1,seesaw2,seesaw3} we use to determine the violation of the inequalities consists of repeating these steps iteratively starting from some initial values until convergence is achieved. This algorithm has been used as well in Ref.~\cite{VB} to get the nonlocal $3\times 3$ bound entangled state.

For each inequality corresponding to $d\leq 8$ we have found ($d\times d$)-dimensional PPT states violating it using the seesaw algorithm. We have not found any such Bell violating PPT state defined in component spaces of less than $d$ dimensions. The maximum violation we have obtained with PPT states is shown in Table~\ref{tab1qviol}.
\begin{table}
\begin{center}\begin{tabular}{cc}
\hline
  $d$ & Quantum violation \\
  \hline
  3 & 0.000265264\\
  4 & 0.000210913\\
  5 & 0.000162725\\
  6 & 0.000128375\\
  7 & 0.000103852\\
  8 & 0.000085873\\
  \hline
\end{tabular}
\caption{Maximum quantum violation for different local dimensions $d$ of the PPT states using see-saw search.}
\label{tab1qviol}
\end{center}
\end{table}
In all cases we have arrived at density matrices invariant to the partial transposition; this is what ensures the PPT property.  Their rank is $2d-2$. For $d>3$ there are two nondegenerate and two ($d-2$)-time-degenerate eigenvalues. From the above observations, we conjecture that the class of Bell inequalities $I_d$, whose $d$th member is defined by the inequality~(\ref{eq:alt1ineq}), gives rise to a dimension witness for bound entangled states: Violation of $I_d$ for $d\ge 3$ using bound entangled states implies that the dimension of the state has to be at least $d\times d$. Note that several recent works presented dimension witnesses of states based on Bell violations placing no additional restrictions on the state (see, e.g., references~\cite{dimwit1,dimwit2,dimwit3,dimwit4,dimwit5,dimwit6}), using as well certain restrictions such as the amount of randomness~\cite{dimwitshared} or the number of singlet pairs~\cite{dimwitmaxent} shared between the parties.

\subsection{Measurement operators}

The eigenvectors of the density matrix belonging to either of the two non-degenerate eigenvalues are such that $d-2$ of their Schmidt coefficients are equal. Moreover, the equal Schmidt coefficients for both eigenstates define the same subspaces of both Alice's and Bob's component spaces.
If we choose the basis vectors labeled with $k=2,\dots,(d-1)$ such that they span these subspaces, the operators for the optimal measurement settings we have got for each $d$ can be written as
\begin{align}
&\hat A_{0|q}=|A_{0|q}\rangle\langle A_{0|q}|,\nonumber\\
&\hat A_{1|q}=\hat I_A-\hat A_{0|q},\nonumber\\
&\hat B_{q|0}=|B_{q|0}\rangle\langle B_{q|0}|,\nonumber\\
&\hat B_{0|1}=|B_{0|1}\rangle\langle B_{0|1}|,\nonumber\\
&\hat B_{1|1}=\hat I_B-\hat B_{0|1},
\label{eq:measopers}
\end{align}
where $q=0,\dots,d-1$, and $\hat I_A$ and $\hat I_B$ are the identity operators in Alice's and Bob's component space, respectively. The vectors appearing on the right-hand side of Eqs.~(\ref{eq:measopers}) may be given with three independent parameters:
\begin{align}
&|A_{0|0}\rangle=|0\rangle_A,\nonumber\\
&|A_{0|p}\rangle=x_0|0\rangle_A+x_1|1\rangle_A+x_2|\theta_{p}\rangle_A,\nonumber\\
&|B_{0|0}\rangle=-y_1|0\rangle_B+y_0|1\rangle_B,\nonumber\\
&|B_{p|0}\rangle=\frac{1}{\sqrt{d-1}}\left(y_0|0\rangle_B+y_1|1\rangle_B+\sqrt{d-2}|\theta_{p}\rangle_B\right),\nonumber\\
&|B_{0|1}\rangle=|0\rangle_B,
\label{eq:measvects}
\end{align}
where $p=1,\dots,d-1$, $x_0^2+x_1^2+x_2^2=1$, $y_0^2+y_1^2=1$, and the vectors $|\theta_{p}\rangle$ are unit vectors in the $(d-2)$-dimensional subspace spanned by $|2\rangle,\dots,|d-1\rangle$ pointing towards the vertices of a regular $d-2$ simplex. They obey the following equations:
\begin{align}
&\sum_{p=1}^{d-1}|\theta_{p}\rangle=0\label{eq:thetrela}\\
&\langle\theta_{p}|\theta_{q}\rangle=\frac{-1+(d-1)\delta_{pq}}{d-2}\label{eq:thetrelb}\\
&\sum_{p=1}^{d-1}|\theta_{p}\rangle\langle\theta_p|=\frac{d-1}{d-2}\sum_{k=2}^{d-1}|k\rangle\langle k|\label{eq:thetrelc}\\
&\sum_{p=1}^{d-1}|\theta_{p},\theta_p\rangle=\frac{d-1}{d-2}\sum_{k=2}^{d-1}|k,k\rangle\label{eq:thetreld},
\end{align}
where $|j,k\rangle$ denotes $|j\rangle_A\otimes|k\rangle_B$. It is important to note that all measurements defined above are of the von Neumann type: The operators giving their components are orthogonal projectors.
For the two-outcome measurements given in Eq.~(\ref{eq:measopers}) this is trivial, and it is easy to check that it is true for Bob's $d$-outcome measurement zero too: Using Eqs.~(\ref{eq:measvects})
and (\ref{eq:thetrelb}) one can see that the $|B_{q|0}\rangle$ ($q=0,\dots,d-1$) vectors are orthonormal.
The measurement operators above are similar to the ones given by Yu and Oh~\cite{YO}. However, they use only one parameter, and their $\theta$ vectors are defined in a ($d-1$)-dimensional space, one dimension larger than ours. Formulas analogous to Eqs.~(\ref{eq:thetrela}-\ref{eq:thetreld}) appear in their paper too.
In their case the angles between all pairs of vectors corresponding to outcome zero of Alice's measurements are the same, that is $\langle A_{0|q}|A_{0|q'}\rangle$ is the same for any $q\neq q'$, including $q=0$. One of our extra parameters breaks this symmetry for Alice's measurement zero. Another symmetry apparent in their case is that
$\langle B_{q|0}|B_{0|1}\rangle$ is the same for any $q$. Our other parameter breaks this symmetry: The value for $q=0$ is not exactly the same as the value for $q\neq 0$. However, for the optimal settings the symmetries are not broken very much. For $d$ up to eight the difference between the violation that we can get using the one-parameter formula and the maximum we could achieve is less than half a percent.

\subsection{Family of $d\times d$ PPT entangled states}

The density operator we have obtained can be parametrized as
\begin{equation}
\hat\rho=\hat S_0+\hat S_1+\hat D_0+\hat D_1,
\label{eq:densmat}
\end{equation}
where
\begin{align}
&\hat S_i=|S_i\rangle\langle S_i|,\label{eq:sop}\\
&\hat D_i=\sum_{k=2}^{d-1}|D_{ik}\rangle\langle D_{ik}|,\label{eq:dop}
\end{align}
with
\begin{align}
&|S_0\rangle=a_{00}|0,0\rangle+a_{01}|0,1\rangle+a_{10}|1,0\rangle+a_{11}|1,1\rangle+A|X\rangle,\label{eq:svec0}\\
&|S_1\rangle=b_{00}|0,0\rangle+b_{01}|0,1\rangle+b_{10}|1,0\rangle+b_{11}|1,1\rangle+B|X\rangle,\label{eq:svec1}\\
&|D_{0k}\rangle=u_0|0,k\rangle+u'_0|k,0\rangle+u_1|1,k\rangle+u'_1|k,1\rangle+U|\varphi_k\rangle,\label{eq:dvec0}\\
&|D_{1k}\rangle=v_0|0,k\rangle+v'_0|k,0\rangle+v_1|1,k\rangle+v'_1|k,1\rangle+V|\varphi_k\rangle,\label{eq:dvec1}
\end{align}
where
\begin{align}
&|X\rangle\equiv\frac{1}{\sqrt{d-2}}\sum_{k=2}^{d-1}|k,k\rangle\label{eq:X}\\
&|\varphi_k\rangle\equiv\frac{(d-2)^{3/2}}{(d-1)\sqrt{d-3}}\sum_{p=1}^{d-1}|\theta_p,\theta_p\rangle\langle\theta_p|k\rangle\label{eq:phik}.
\end{align}
All parameters in the equations above are real numbers, as we have never got any larger violation when we allowed complex numbers in the numerical optimization.
The $|\varphi_k\rangle$ vectors appear in the construction of the density matrix of Yu and Oh too, but in their case they are defined in a ($d-1$)-dimensional space~\cite{YO}. The vectors are orthonormal, as it can be checked using Eqs.~(\ref{eq:thetrela}-\ref{eq:thetrelc}). Using the same relations one can also show that
$|\varphi_k\rangle$ and the unit vector $|X\rangle$ are orthogonal. From these it follows that
$\langle D_{ik}|S_{j}\rangle=0$ and $\langle D_{ik}|D_{jk'}\rangle=0$ if $k\neq k'$. The vectors in
Eqs.~(\ref{eq:svec0}-\ref{eq:dvec1}) are not normalized, the square of their norm is the probability associated with them. One constraint the parameters must obey is that the sum of all probabilities is one, that is
\begin{align}
&p^S_0=a_{00}^2+a_{01}^2+a_{10}^2+a_{11}^2+A^2,\nonumber\\
&p^S_1=b_{00}^2+b_{01}^2+b_{10}^2+b_{11}^2+B^2,\nonumber\\
&p^D_0=u_0^2+u^{\prime 2}_0+u_1^2+u^{\prime 2}_1+U^2,\nonumber\\
&p^D_1=v_0^2+v^{\prime 2}_0+v_1^2+v^{\prime 2}_1+V^2,\nonumber\\
&p^S_0+p^S_1+(d-2)(p^D_0+p^D_1)=1.
\label{eq:prob}
\end{align}
A freedom we have in choosing the parameters is that the transformations
$|S'_i\rangle=\sum_{j=0}^1O_{ij}^S|S_j\rangle$ and $|D'_{ik}\rangle=\sum_{j=0}^1O_{ij}^D|D_{jk}\rangle$,
where $O_{ij}^S$ and $O_{ij}^D$ are $2\times 2$ orthogonal matrices, leave both the density matrix of Eq.~(\ref{eq:densmat}) and the form of Eqs.~(\ref{eq:svec0}-\ref{eq:dvec1}) unchanged.
Using this freedom we can ensure that all vectors are pairwise orthogonal. These vectors will be eigenvectors of the density matrix and the squares of their norms, that is the probabilities in Eq.~(\ref{eq:prob}) will be its eigenvalues. There will
be two non-degenerate eigenvalues corresponding to the $S$-vectors and two $(d-2)$-times degenerated ones corresponding to the $D$-vectors, as we have already stated. Later it will be more convenient for us to use our freedom not to make all vectors orthogonal, but to eliminate parameters $B$ and $V$.

Besides Eq.~(\ref{eq:prob}) the parameters obey further constraints to ensure invariance to partial transposition. These are the following:
\begin{align}
&A^2+B^2=\frac{d-2}{d-3}(U^2+V^2),\label{eq:invcons1}\\
&a_{00}a_{11}+b_{00}b_{11}=a_{01}a_{10}+b_{01}b_{10},\label{eq:invcons2}\\
&Aa_{00}+Bb_{00}=\sqrt{d-2}(u_0u'_0+v_0v'_0),\label{eq:invcons3}\\
&Aa_{01}+Bb_{01}=\sqrt{d-2}(u_0u'_1+v_0v'_1),\label{eq:invcons4}\\
&Aa_{10}+Bb_{10}=\sqrt{d-2}(u_1u'_0+v_1v'_0),\label{eq:invcons5}\\
&Aa_{11}+Bb_{11}=\sqrt{d-2}(u_1u'_1+v_1v'_1).\label{eq:invcons6}
\end{align}
The proof is given in Appendix~\ref{apppartran}.

\subsection{Analytic expressions for the violation}\label{analyt}

The quantum value in Eqs.~(\ref{eq:qtrrB}-\ref{bellopa})
with the density operator of Eqs.~(\ref{eq:densmat}-\ref{eq:phik}) and measurement setting according to Eqs.~(\ref{eq:measopers}-\ref{eq:measvects})
may be calculated analytically. The details of the calculation are given in Appendix~\ref{appa}. The result can be written as
\begin{equation}
Q(d)=Q_{S_0}(d)+Q_{S_1}(d)+Q_{D_0}(d)+Q_{D_1}(d),
\label{eq:violat}
\end{equation}
where $Q_{S_0}(d)$ and $Q_{D_0}(d)$ are given in Eq.~(\ref{eq:S0contrib}) and Eq.~(\ref{eq:D0contrib}), respectively,
while $Q_{S_1}(d)$ and $Q_{D_1}(d)$ have the same form as $Q_{S_0}(d)$ and $Q_{D_0}(d)$ but with other parameters, as explained in Appendix~\ref{appa}.

\begin{figure}[ht]
\vspace{0.1cm}
\centering
\includegraphics[angle=-90, width=0.95\columnwidth]{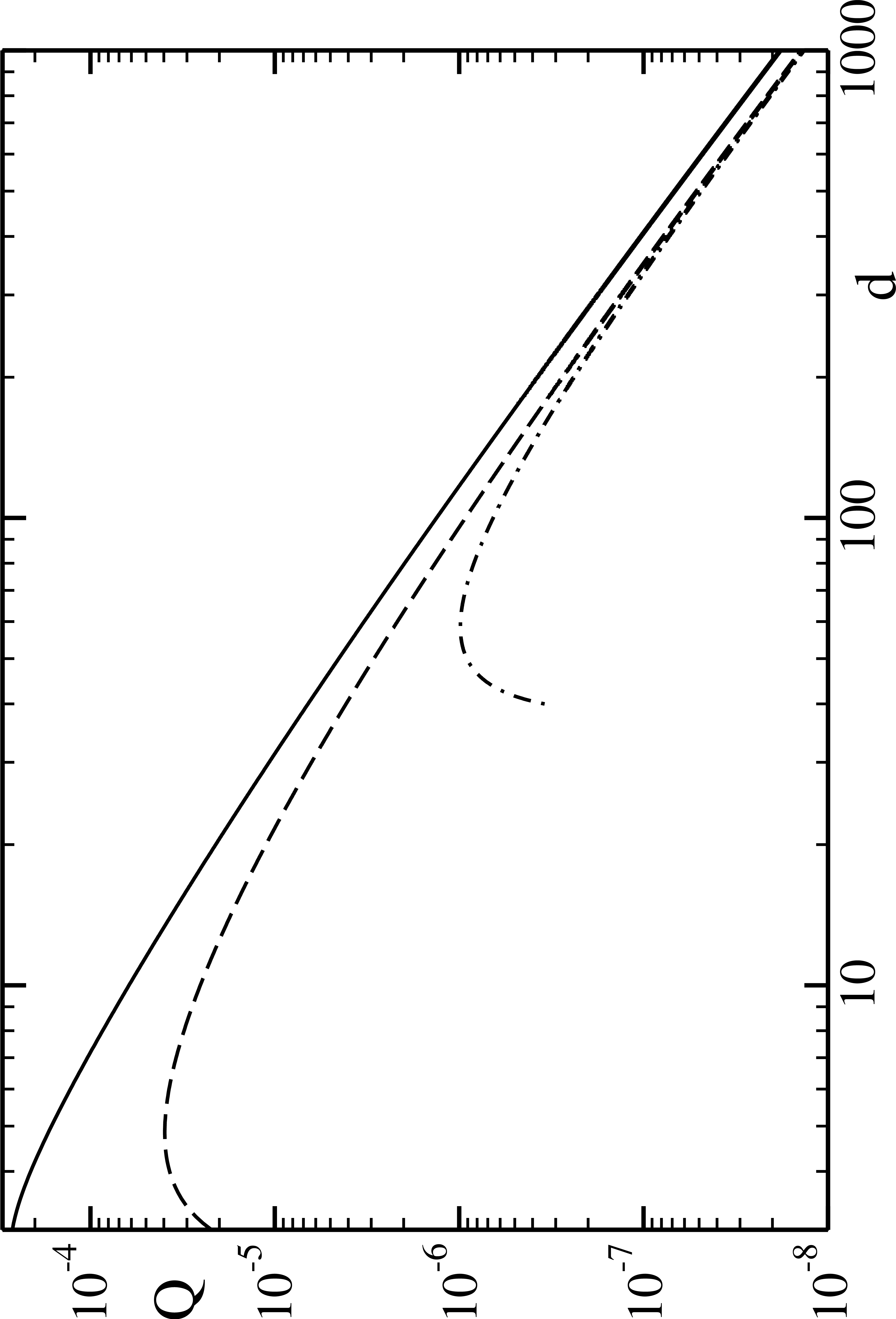}
\caption{Violation $Q$ of the family of Bell inequalities of Eq.~(\ref{eq:alt1ineq}). Solid line: full set of parameters. Dashed line: result of a suboptimal solution with a reduced number of parameters, as given by the values~(\ref{eq:zerpars}). Dash-dotted line: the suboptimal solution with asymptotic parameter values, as given in Eq.~(\ref{eq:Qdapprnum}).}
\label{fig:viola}
\end{figure}

There are altogether 20 parameters in Eqs.~(\ref{eq:svec0}-\ref{eq:dvec1}) defining the density matrix. Due to the normalization condition given in Eq.~(\ref{eq:prob}) and our freedom of choice we noted under Eq.~(\ref{eq:prob}) the number of free parameters is reduced by three. We use our freedom to take $B=V=0$. Taking into account the six constraints given in Eqs.~(\ref{eq:invcons1}-\ref{eq:invcons6}), eleven free parameters remain to
determine the density matrix. Three more independent parameters are necessary for the measurement settings (see Eq.~(\ref{eq:measvects})). We have used an uphill simplex method~\cite{amoeba} with 14 parameters to find the optimum violation. In Fig.~\ref{fig:viola}, solid line, we show the result of this calculation up to $d=1000$.

The function reaches its asymptotic behavior very slowly; the log-log plot is not quite straight even around $d=1000$. There its slope is compatible with a function proportional to $d^{-1.9}$. Unfortunately, the number of parameters has been too large to allow us to give an explicit analytical solution, even asymptotically. We can get a suboptimal solution by choosing five parameters (besides $B$ and $V$) to be zero, while still getting a violation of the Bell inequality (a sixth parameter also becomes zero due to a constraint). The parameters of value zero are given in Eq.~(\ref{eq:zerpars}) in Appendix~\ref{appredu}. We have also chosen Bob's measurement settings to be parameter-free as Yu and Oh~\cite{YO}. We have kept the extra parameter we have introduced for Alice's settings. The violation coming from this suboptimal solution is shown in Fig.~(\ref{fig:viola}) by a dashed line. For small $d$ it gives a much smaller violation than the optimum one with the full set of parameters, but for $d=1000$ the difference becomes much smaller, about 23\%. Unfortunately, this suboptimal solution will almost certainly not converge to the optimal one at the $d=\infty$ limit. However, its asymptotic behavior can be determined analytically. The details are given in the Appendix~\ref{appredu}. The violation with the asymptotic parameter values is given explicitly in Eq.~(\ref{eq:Qdapprnum}) and it is shown in Fig.~(\ref{fig:viola}) by a dash-dotted line. The leading order term behaves as $d^{-2}$, which is to be compared to the $d^{-4}$ scaling of the family of Yu and Oh~\cite{YO}. Furthermore, there is also a term proportional to $d^{-5/2}$, with a factor more than six times larger than that of the leading order term, which explains why convergence to the asymptotic behavior is so slow.

\section{Summary}
\label{Disc}

We have studied the violation of a family of Bell inequalities by bound entangled states. We have shown that these inequalities may be violated by such states. Each inequality can be characterized by an integer $d\geq 3$ and it corresponds to a bipartite Bell scenario with $d$ two-outcome measurement settings for one party and one $d$-outcome and one two-outcome measurement setting for the other one.  The family is the generalization of one of the inequalities we have got when we have constructed all tight inequalities with $d=4$.

To find PPT states and measurement operators violating the inequalities we have used a numerical method, a see-saw algorithm for $d$ values up to eight. Such an algorithm does not guarantee the optimum solution even when repeated several times starting from different initial values. Nevertheless, when it has found a solution, for each $d$ most of the time it has found the same one up to transformations of the local coordinate systems. The corresponding state may be given in a ($d\times d$)-dimensional Hilbert space. Our experience with the seesaw method is that if solutions of different dimensionalities violating the inequality exist, the algorithm will find the lowest-dimensional one with highest probability. None of our attempts has led to a solution of less than $d\times d$ dimensions. Moreover, we have tried to find a solution while we explicitly restricted the search to lower-dimensional spaces and we failed to find any. We are quite confident that at least for $d=4$ and for $d=5$ we would have found such a solution if it existed. Based on these numerical experiences we conjecture that the Bell inequality characterized by $d$ belonging to this family may only be violated by bound entangled states of dimensions of at least $d\times d$, that is, it acts as a dimension witness for bound entangled states. This is an important difference between our family and the one proposed by Yu and Oh~\cite{YO}, whose members may all equally be violated by the $d=3$ solution.

All the optimal solutions we have found numerically for up to $d=8$ have a certain well-defined mathematical form when the appropriate local bases are chosen in the component Hilbert spaces. We have used this form as an ansatz to find solutions violating the inequalities up to $d=1000$. We have also given a simplified suboptimal solution in a fully analytic form still violating the inequalities.

The Bell violation of the inequalities with the constructed PPT states tends to zero as $d$ goes to infinity. From the analytic suboptimal solution we can conclude that the violation decreases no faster than $d^{-2}$. It remains to be seen if other type of bipartite Bell inequalities using the same family of states may lead to increased Bell violation with increasing $d$.  Higher violation implies in general higher noise resistance, hence this property would be useful in certain quantum information tasks based on nonlocality, such as communication complexity problems~\cite{zukowski,epping}. Positive partial transpose states are known to be useful in quantum key distribution~\cite{hllo,ozols}, and it is an interesting question if they exhibit a private key in the device-independent scenario as well~\cite{karol}. Since entanglement and steerability are necessary ingredients to Bell nonlocality, it will also be interesting to look at suitable entanglement witnesses~\cite{EW1,EW2,EW3} or steerability witnesses~\cite{SW1,SW2} associated with our states similarly to the states in Ref.~\cite{YO}.

\section{Acknowledgements} We thank N. Brunner, F. Hirsch, and M.T. Quintino for useful discussions, and J. Kaniewski for pointing out a mistake in one of the formulas for the rank of the state. We acknowledge financial support from the Hungarian National Research Fund OTKA (No.~K111734 and No.~KH125096).





\appendix
\onecolumngrid
\section{Partial transposition invariance}
\label{apppartran}

Here we show that the constraints given in Eqs.~(\ref{eq:invcons1}-\ref{eq:invcons6}) ensure that the density matrix of Eq.~(\ref{eq:densmat}) is invariant to partial transposition.
From Eqs.~(\ref{eq:sop}) and (\ref{eq:svec0}) we get
\begin{equation}
\hat S_0=\hat S_0^{inv}+\hat S_0^{SS}+\hat S_0^{SD},
\label{eq:S0parts}
\end{equation}
where
\begin{align}
\hat S_0^{inv}=&a_{00}^2|0,0\rangle\langle 0,0|+a_{01}^2|0,1\rangle\langle 0,1|+a_{10}^2|1,0\rangle\langle 1,0|+a_{11}^2|1,1\rangle\langle 1,1|\nonumber\\
&+a_{00}a_{01}(|0,0\rangle\langle 0,1|+|0,1\rangle\langle 0,0|)+a_{00}a_{10}(|0,0\rangle\langle 1,0|+|1,0\rangle\langle 0,0|)\nonumber\\
&+a_{01}a_{11}(|0,1\rangle\langle 1,1|+|1,1\rangle\langle 0,1|)+a_{10}a_{11}(|1,0\rangle\langle 1,1|+|1,1\rangle\langle 1,0|)\label{eq:S0inv}\\
\hat S_0^{SS}=&a_{00}a_{11}(|0,0\rangle\langle 1,1|+|1,1\rangle\langle 0,0|)+a_{01}a_{10}(|0,1\rangle\langle 1,0|+|1,0\rangle\langle 0,1|)\label{eq:S0SS}\\
\hat S_0^{SD}=&\frac{A}{\sqrt{d-2}}\sum_{k=2}^{d-1}[a_{00}(|0,0\rangle\langle k,k|+|k,k\rangle\langle 0,0|)+a_{01}(|0,1\rangle\langle k,k|+|k,k\rangle\langle 0,1|)\nonumber\\
&+a_{10}(|1,0\rangle\langle k,k|+|k,k\rangle\langle 1,0|)+a_{11}(|1,1\rangle\langle k,k|+|k,k\rangle\langle 1,1|)]+A^2|X\rangle\langle X|.
\label{eq:S0SD}
\end{align}
For Eq.~(\ref{eq:S0SD}) we have also used Eq.~(\ref{eq:X}). From Eqs.~(\ref{eq:sop}) and (\ref{eq:svec1}) we can get the analogous expressions for $\hat S_1$ with $A$ and $a_{ij}$ are replaced by $B$ and $b_{ij}$, respectively. It is easy to see that $\hat S_0^{inv}$ and $\hat S_1^{inv}$ are invariant to partial transposition for any values of the parameters. The operators $(|0,0\rangle\langle 1,1|+|1,1\rangle\langle 0,0|)$ and $(|0,1\rangle\langle 1,0|+|1,0\rangle\langle 0,1|)$ appearing in $\hat S_0^{SS}$ and $\hat S_1^{SS}$ are the partial transpose of each other. Their factors have to be equal for the invariance, which is just the constraint of Eq.~(\ref{eq:invcons2}) ($\hat D_i$ contains no such terms).

From Eqs.~(\ref{eq:dop}) and (\ref{eq:dvec0}) we get
\begin{equation}
\hat D_0=\hat D_0^{inv}+\hat D_0^{DS},
\label{eq:D0parts}
\end{equation}
where
\begin{align}
\hat D_0^{inv}=&\sum_{k=2}^{d-1}[u_0^2|0,k\rangle\langle 0,k|+u^{\prime 2}_0|k,0\rangle\langle k,0|+u_1^2|1,k\rangle\langle 1,k|+u^{\prime 2}_1|k,1\rangle\langle k,1|\nonumber\\
&+u_0u_1(|0,k\rangle\langle 1,k|+|1,k\rangle\langle 0,k|)+u'_0u'_1(|k,0\rangle\langle k,1|+|k,1\rangle\langle k,0|)\nonumber\\
&+u_0U(|0,k\rangle\langle\varphi_k|+|\varphi_k\rangle\langle 0,k|)+u'_0U(|k,0\rangle\langle\varphi_k|+|\varphi_k\rangle\langle k,0|)\nonumber\\
&+u_1U(|1,k\rangle\langle\varphi_k|+|\varphi_k\rangle\langle 1,k|)+u'_1U(|k,1\rangle\langle\varphi_k|+|\varphi_k\rangle\langle k,1|)]\label{eq:D0inv}\\
\hat D_0^{DS}=&\sum_{k=2}^{d-1}[u_0u'_0(|0,k\rangle\langle k,0|+|k,0\rangle\langle 0,k|)+u_0u'_1(|0,k\rangle\langle k,1|+|k,1\rangle\langle 0,k|)\nonumber\\
&+u_1u'_0(|1,k\rangle\langle k,0|+|k,0\rangle\langle 1,k|)+u_1u'_1(|1,k\rangle\langle k,1|+|k,1\rangle\langle 1,k|)+U^2|\varphi_k\rangle\langle\varphi_k|].\label{eq:D0DS}
\end{align}
To get $\hat D_1$ and its parts one should only replace $u_i$, $u'_i$ and $U$ for $v_i$, $v'_i$ and $V$, respectively. Parts $\hat D_0^{inv}$ and
$\hat D_1^{inv}$ are invariant to partial transposition. For terms containing no $|\varphi_k\rangle$ or $\langle\varphi_k|$ this is obvious. As far as the rest of the terms are concerned, let us take
 \begin{align}
 \sum_{k=2}^{d-1}(|0,k\rangle\langle\varphi_k|+|\varphi_k\rangle\langle 0,k|)=&\frac{(d-2)^{3/2}}{(d-1)\sqrt{d-3}}\sum_{p=1}^{d-1}\sum_{k=2}^{d-1}(|0,k\rangle\langle k|\theta_p\rangle\langle\theta_p,\theta_p\rangle+|\theta_p,\theta_p\rangle\langle\theta_p|k\rangle\langle 0,k|)\nonumber\\
 =&\frac{(d-2)^{3/2}}{(d-1)\sqrt{d-3}}\sum_{p=1}^{d-1}(|0,\theta_p\rangle\langle\theta_p,\theta_p|+|\theta_p,\theta_p\rangle\langle 0,\theta_p|),
 \label{eq:invphipart}
\end{align}
which is also invariant. Here we have used Eq.~(\ref{eq:phik}) and the fact that each $|\theta_p\rangle$ is in the subspace spanned by $|k\rangle$ ($2\leq k\leq d-1$). The invariance of the other terms can be shown similarly.

The operators $(|i,j\rangle\langle k,k|+|k,k\rangle\langle i,j|)$ ($i,j=0,1$ and $2\leq k\leq d-1$) appearing in $\hat S_0^{SD}$ and $\hat S_1^{SD}$ and $(|i,k\rangle\langle k,j|+|k,j\rangle\langle i,k|)$ appearing in $\hat D_0^{DS}$ and $\hat D_1^{DS}$ are the partial transpose of each other. The equality of their factors can be ensured by the constraints Eqs.~(\ref{eq:invcons3}-\ref{eq:invcons6}).

The operator in the last term remaining in $\hat D_0^{DS}$ and $\hat D_1^{DS}$ may be rewritten as:
\begin{equation}
\sum_{k=2}^{d-1}|\varphi_k\rangle\langle\varphi_k|=\frac{(d-2)^2}{(d-1)(d-3)}\sum_{p=1}^{d-1}|\theta_p,\theta_p\rangle\langle\theta_p,\theta_p|-\frac{d-2}{d-3}|X\rangle\langle X|,
\label{eq:phiiden}
\end{equation}
which can be proven by using Eqs.~(\ref{eq:phik}), (\ref{eq:thetrelb}-\ref{eq:thetreld}) and (\ref{eq:X}).
The sum on the right-hand side is invariant to partial transposition, but $|X\rangle\langle X|$ is not. However, such a term also appears in $\hat S_0^{SD}$ and $\hat S_1^{SD}$. If Eq.~(\ref{eq:invcons1}) holds, this noninvariant term is eliminated.

\section{Details of the calculation of the Bell value}
\label{appa}

The measurement operators we are going to consider here are the
ones given in Eqs.~(\ref{eq:measopers})-(\ref{eq:measvects}). We
may rewrite the Bell operator in Eq.~(\ref{bellopa}) as
\begin{equation}
\hat{\cal B}_d=(d-2)\hat A_{0|0}\otimes(\hat B_{0|1}-\hat B_{0|0})-\sum_{i,j=1}^{d-1}\hat A_{0|i}\otimes\hat B_{j|0}(1-\delta_{ij})-\sum_{i=1}^{d-1}(\hat I_A-\hat A_{0|i})\otimes\hat B_{0|1}.
\label{bellop}
\end{equation}
We have used $\hat A_{1|i}=\hat I_A-\hat A_{0|i}$ to ensure that all measurement operators appearing in
the expression above are one-dimensional projectors.

Our density operator of Eqs.~(\ref{eq:densmat}), (\ref{eq:sop}) and (\ref{eq:dop}) is written as a sum of terms
of the form $\hat T=|T\rangle\langle T|$, where $\hat T$ is either $\hat S_i$ or $\hat D_{ik}$. The contribution of each term to a conditional probability in Eq~(\ref{eq:condprob}) may be written as
\begin{equation}
\Tr[\hat T(\hat A_{a|x}\otimes\hat B_{b|y})]=\langle T|\hat A_{a|x}\otimes\hat B_{b|y}|T\rangle.
\label{eq:probterms}
\end{equation}
If the measurement operators are
one-dimensional projectors, then Eq.~(\ref{eq:probterms}) may further be simplified as:
\begin{equation}
\Tr[\hat T(\hat A_{a|x}\otimes\hat B_{b|y})]=|\langle A_{a|x},B_{b|y}|T\rangle|^2.
\label{eq:probtermss}
\end{equation}
Now let us calculate the contribution of the first term $\hat S_0$ of the density matrix to the quantum value. Using Eqs.~(\ref{eq:svec0}), (\ref{eq:X}) and (\ref{eq:thetreld}) we can get
\begin{equation}
|S_0\rangle=\sum_{\alpha,\beta=0}^1 a_{\alpha\beta}|\alpha,\beta\rangle+A\frac{\sqrt{d-2}}{d-1}\sum_{p=1}^{d-1}|\theta_p,\theta_p\rangle.
\label{eq:s0app}
\end{equation}
Then using Eqs.~(\ref{eq:measvects}) and the identity
\begin{equation}
\langle\theta_i,\theta_j|S_0\rangle=\frac{A}{(d-2)^{3/2}}[(d-1)\delta_{ij}-1],
\end{equation}
which one can get by using Eqs.~(\ref{eq:s0app}) and (\ref{eq:thetrelb}), we arrive at
\begin{align}
|\langle A_{0|0},B_{0|1}|S_0\rangle|^2=&a_{00}^2\label{eq:Sterms1}\\
|\langle A_{0|0},B_{0|0}|S_0\rangle|^2=&(-y_1a_{00}+y_0a_{01})^2\label{eq:Sterms2}\\
|\langle A_{0|i},B_{j|0}|S_0\rangle|^2=&\frac{1}{d-1}\Bigg[\sum_{\alpha,\beta=0}^1x_{\alpha}y_{\beta}a_{\alpha\beta}-\frac{x_2A}{d-2}[1-(d-2)\delta_{ij}]\Bigg]^2\label{eq:Sterms3}\\
|\langle A_{0|i},B_{0|1}|S_0\rangle|^2=&(x_0a_{00}+x_1a_{10})^2,\label{eq:Sterms4}
\end{align}
where $i,j\geq 1$, and also we get
\begin{equation}
\langle S_0|\hat I_A\otimes\hat B_{0|1}|S_0\rangle=a_{00}^2+a_{10}^2.
\label{eq:Sterms5}
\end{equation}
We note that the right-hand side of Eq.~(\ref{eq:Sterms3}) is independent of indices $i$ and $j$ if $i\neq j$ and the right-hand side of Eq.~(\ref{eq:Sterms4}) is independent of $i$. Then the contribution to the quantum
value coming from $\hat S_0$ is
\begin{equation}
Q_{S_0}(d)=(d-2)[a^2_{00}-(y_0a_{01}-y_1a_{00})^2]-
(d-2)\left[\sum_{\alpha,\beta=0}^1x_{\alpha}y_{\beta}a_{\alpha\beta}-\frac{x_2A}{d-2}\right]^2+
(d-1)[(x_0a_{00}+x_1a_{10})^2-a^2_{00}-a^2_{10}].
\label{eq:S0contrib}
\end{equation}
The form of the contribution $Q_{S_1}(d)$ from $\hat S_1$ is the same, only $a_{\alpha\beta}$ and $A$ should be replaced by
$b_{\alpha\beta}$ and $B$, respectively.

Now let us calculate the contribution from $\hat D_0$. From Eqs.~(\ref{eq:dvec0}) and (\ref{eq:measvects}) it is easy to see that
$\langle A_{0|0},B_{0|1}|D_{0k}\rangle=\langle A_{0|0},B_{0|0}|D_{0k}\rangle=0$.
To calculate $\langle A_{0|i},B_{j|0}|D_{0k}\rangle$ we also need the identity
\begin{equation}
\langle\theta_i,\theta_j|D_{0k}\rangle=-\frac{U}{\sqrt{(d-2)(d-3)}}[\langle\theta_i|k\rangle+\langle\theta_i|k\rangle-(d-1)\delta_{ij}\langle\theta_i|k\rangle],
\label{identforD}
\end{equation}
which may be derived from Eqs.~(\ref{eq:dvec0}), (\ref{eq:phik}), (\ref{eq:thetrelb}), and (\ref{eq:thetrela}). Then, if $i,j\geq 1$ and $i\neq j$ we get
\begin{equation}
|\langle A_{0|i},B_{j|0}|D_{0k}\rangle|^2=\frac{1}{d-1}|F_0\langle\theta_j|k\rangle+G_0\langle\theta_i|k\rangle|^2,
\label{eq:Dkterms3}
\end{equation}
where
\begin{align}
&F_0\equiv\sqrt{d-2}(x_0u_0+x_1u_1)-\frac{x_2U}{\sqrt{d-3}}\label{eq:F0}\\
&G_0\equiv x_2(y_0u'_0+y_1u'_1)-\frac{x_2U}{\sqrt{d-3}}\label{eq:G0}.
\end{align}
As vectors $|\theta_i\rangle$ are in the subspace spanned by vectors $|2\rangle,\dots,|d-1\rangle$,
\begin{equation}
\sum_{k=2}^{d-1}|\langle A_{0|i},B_{j|0}|D_{0k}\rangle|^2=\frac{1}{d-1}\left(F^2_{0}+G^2_{0}-\frac{2}{d-2}F_0G_0\right),
\label{eq:Dterms3}
\end{equation}
which is independent of $i$ and $j$. Similarly, for the last terms needed one can get
\begin{equation}
\sum_{k=2}^{d-1}|\langle A_{0|i},B_{0|1}|D_{0k}\rangle|^2=\sum_{k=2}^{d-1}|x_2u'_0\langle\theta_i|k\rangle|^2=x_2^2u^{\prime 2}_0,
\label{eq:Dterms4}
\end{equation}
and
\begin{equation}
\sum_{k=2}^{d-1}\langle D_{0k}|\hat I_A\otimes\hat B_{0|1}|D_{0k}\rangle=(d-2)u^{\prime 2}_0.
\label{eq:Dterms5}
\end{equation}
Putting the terms above together, one gets the contribution from $\hat D_0$:
\begin{equation}
Q_{D_0}(d)=2F_0G_0-(d-2)(F^2_{0}+G^2_{0})-(d-1)u^{\prime 2}_0(d-2-x_2^2).
\label{eq:D0contrib}
\end{equation}
The contribution from $\hat D_1$ has the same form, only $F_0$, $G_0$, and $u'_0$ should be replaced by $F_1$, $G_1$, and $v'_0$ , respectively, where $F_1$ and $G_1$ are defined like $F_0$ and $G_0$ in Eqs.~(\ref{eq:F0}) and (\ref{eq:G0}), only $u_\mu$ and $u'_\mu$ are replaced by $v_\mu$ and $v'_\mu$, respectively.  We note that these contributions can not be positive if $d\geq 3$.

\section{A reduced number of parameters}\label{appredu}

We may take all the following parameters to be zero and still get a violation of the Bell inequality:
\begin{equation}
a_{01}=b_{01}=b_{10}=B=u_0=u'_0=v'_1=V=0.
\label{eq:zerpars}
\end{equation}
Although with this choice the violation is 88\% smaller for $d=3$ and 77\% smaller for $d=4$ than before, for $d=1000$ the violation is only reduced by about 15\% if the remaining parameters are optimally chosen. To keep partial transposition invariance according to Eqs.~(\ref{eq:invcons1}-\ref{eq:invcons6}), the following relations must hold:
\begin{align}
&A=\sqrt{\frac{d-2}{d-3}}U\label{eq:invconsred1}\\
&a_{00}a_{11}+b_{00}b_{11}=0\label{eq:invconsred2}\\
&Aa_{00}=\sqrt{d-2}v_0v'_0\label{eq:invconsred3}\\
&Aa_{10}=\sqrt{d-2}v_1v'_0\label{eq:invconsred5}\\
&Aa_{11}=\sqrt{d-2}u_1u'_1.\label{eq:invconsred6}
\end{align}
Eq.~(\ref{eq:invcons4}) is automatically satisfied with both of its sides zero. The relations above fix the values of $A$, $a_{00}$, $a_{10}$, $a_{11}$ and $b_{00}b_{11}$ in terms of the other parameters.

Another simplification we have made is that for Bob we used the measurement settings of Yu and Oh~\cite{YO} by choosing the following values in Eq.~(\ref{eq:measvects}):
\begin{equation}
y_0=\sqrt{\frac{d-1}{d}}, \quad y_1=-\frac{1}{\sqrt{d}}.\label{eq:yYO}
\end{equation}
It is easy to check that $\langle B_0^q|B_1^0\rangle=1/\sqrt{d}$ hold for all $q$; therefore, the symmetry for Bob mentioned earlier is valid. This choice decreases the violation somewhat further, for $d=1000$ it is about 77\% of the original one. For Alice we keep both of our independent parameters and will take $x_0>0$. All $x_i$ ($i=0,1,2$) converge to numbers different from zero in the infinite $d$ limit, therefore, unlike in the case discussed by Yu and Oh, Alice's measurements remain distinguishable.

Numerical results show that for $Q_{D_0}(d)$ [see Eq.~(\ref{eq:D0contrib})], the contribution from $\hat D_0$, is always orders of magnitude smaller than the contribution from other terms. The violation changes very marginally if we take this contribution to be zero by demanding $F_0=G_0=0$ ($u'_0=0$, anyway). These requirements will fix two more parameters according to Eqs.~(\ref{eq:F0}), (\ref{eq:G0}) and (\ref{eq:zerpars}) as:
\begin{align}
u_1&=\frac{x_2U}{x_1\sqrt{(d-2)(d-3)}}\nonumber\\
u'_1&=-U\sqrt{\frac{d}{d-3}}.
\label{eq:u1u1p}
\end{align}
Here we used the values of $y_0$ and $y_1$ from Eq.~(\ref{eq:yYO}). The value of $F_1$ can also be taken to be exactly zero with hardly any change of the violation. Then, from the equation analogous with Eq.~(\ref{eq:F0}) [see remark below Eq.~(\ref{eq:D0contrib})], and taking into account that $V=0$, we get the following relation between $v_0$ and $v_1$:
\begin{equation}
x_0v_0+x_1v_1=0.
\label{eq:v0v1}
\end{equation}
Using the equations analogous to Eqs.~(\ref{eq:D0contrib}) and (\ref{eq:G0}) we get for the contribution from $\hat D_1$:
\begin{equation}
Q_{D_1}(d)=-(d-1)(d-2)\left(1-\frac{2x_1^2}{d(d-2)}\right){v'}_0^2.
\label{eq:qd1}
\end{equation}
We have taken into account that $u'_1=0$ and $y_0=\sqrt{(d-1)/d}$.

From Eqs.~(\ref{eq:invconsred6}), (\ref{eq:u1u1p}), (\ref{eq:invconsred1}), and (\ref{eq:yYO}) it follows that
\begin{equation}
x_1y_1a_{11}-\frac{x_2A}{d-2}=0.
\label{eq:a11}
\end{equation}
From Eqs.~(\ref{eq:invconsred3}), (\ref{eq:invconsred5}), and (\ref{eq:v0v1}) it also follows that
\begin{equation}
x_0a_{00}+x_1a_{10}=0.
\label{eq:x0a00x0a10}
\end{equation}
Using Eqs.~(\ref{eq:a11}), (\ref{eq:x0a00x0a10}), (\ref{eq:yYO}), (\ref{eq:invconsred1}), and (\ref{eq:invconsred3}) the contribution of $\hat S_0$ given by Eq.~(\ref{eq:S0contrib}) may be written as
\begin{equation}
Q_{S_0}(d)=-(d-1)(d-3)\left(\frac{x_0^2}{x_1^2}+\frac{2}{d}\right)\frac{v_0^2v_0^{\prime 2}}{U^2}.
\label{eq:qs0}
\end{equation}
The contribution of $\hat S_1$ given by Eq~(\ref{eq:S0contrib}) with $a_{\alpha\beta}$ and $A$ replaced by
$b_{\alpha\beta}$ and $B$, respectively, with $b_{01}=b_{10}=B=0$ and $y_0$ and $y_1$ taken from Eq.~(\ref{eq:yYO}) is simplified as:
\begin{equation}
Q_{S_1}(d)=-2\frac{d-1}{d}(1-x_0^2)b_{00}^2-\frac{d-2}{d}x_1^2b_{11}^2+2\frac{d-2}{d}\sqrt{d-1}x_0x_1b_{00}b_{11},
\label{eq:qs1a}
\end{equation}
As the first two terms are negative, the absolute value of their contribution must be as small as possible to get the maximum violation. This is achieved if the ratio of $b_{00}$ and $b_{11}$ is such that these negative terms are equal, since their product is fixed. This is because the product of $b_{00}$ and $b_{11}$ is fixed by the other parameters according to Eqs.~(\ref{eq:invconsred1}-\ref{eq:invconsred6}). Then we get for the ratio of $b_{00}$ and $b_{11}$:
\begin{equation}
\frac{b_{00}}{b_{11}}=\sqrt{\frac{d-2}{2(d-1)}}\frac{x_1}{\sqrt{1-x_0^2}}
\label{eq:b00pb11}
\end{equation}
The sign of the expression is such that the third term of Eq.~(\ref{eq:qs1a}) is positive. It must be so to get a violation, as all other non-zero terms of all contributions are negative. Using this formula, we can rewrite Eq.~(\ref{eq:qs1a}) as
\begin{equation}
Q_{S_1}(d)=2\frac{d-2}{d}\sqrt{d-1}\left(1-\sqrt{\frac{2(1-x_0^2)}{d-2}}\frac{1}{x_0}\right)\cdot x_0x_1b_{00}b_{11}.
\label{eq:qsb}
\end{equation}
From Eqs.~(\ref{eq:invconsred1}-\ref{eq:invconsred6}) and Eq.~(\ref{eq:u1u1p}) it follows that
$x_1b_{00}b_{11}=x_2\sqrt{d/(d-2)}v_0v'_0$; therefore,
\begin{equation}
Q_{S_1}(d)=2\sqrt{\frac{(d-1)(d-2)}{d}}\left(1-\sqrt{\frac{2(1-x_0^2)}{d-2}}\frac{1}{x_0}\right)\cdot x_0x_2v_0v'_0.
\label{eq:qs1}
\end{equation}

The quantum value $Q(d)$ is the sum of the contributions given by Eqs.~(\ref{eq:qd1}), (\ref{eq:qs0}) and (\ref{eq:qs1}), while $Q_{D_0}(d)=0$. This is the value to be maximized. Let us choose the value of $v'_0$ such that the partial derivative of $Q(d)$ in terms of $v'_0$ is zero. With this choice we get the relationship
\begin{equation}
2Q_{D_0}(d)+2Q_{S_0}(d)+Q_{S_1}(d)=0,
\label{eq:qrelsforvp0}
\end{equation}
from which it follows that
\begin{equation}
Q(d)=\frac{1}{2}Q_{S_1}(d).
\label{eq:qhalfqs1}
\end{equation}
We note that this choice for $v'_0$ is not exactly the optimal one, as the value of $v'_0$ affects the other parameters through the normalization condition given by Eq.~(\ref{eq:prob}). Nevertheless, for large enough $d$ this influence becomes negligible, because $v'_0$ itself becomes negligible compared to the terms dominating the normalization condition. This follows from $|Q_{S_1}(d)|>|Q_{D_0}(d)|$, which must hold to get a positive violation. At the same time $v_0v'_0$ should converge to zero as slowly as possible at the infinite $d$ limit. This is achieved if $v'_0/v_0\propto d^{-3/2}$ for large $d$, and $v_0$ is one of the dominant parameters.
Then it follows from $|Q_{S_1}(d)|>|Q_{S_0}(d)|$ that $U$ must be of the same order as $v_0$. There are just two more parameters that can not be neglected when writing up the normalization condition at the infinite $d$ limit, namely, $u'_1$ [see Eq.~(\ref{eq:u1u1p})] and $v_1$ [see Eq.~(\ref{eq:v0v1})]. For large $d$ these considerations and Eq.~(\ref{eq:prob}) lead approximately to
\begin{align}
d(u^{\prime 2}_1+U^2+v_0^2+v_1^2)&\approx 1\nonumber\\
2U^2+\frac{x_0^2+x_1^2}{x_1^2}v_0^2&\approx 1/d.
\label{eq:normapp}
\end{align}
From now on let us concentrate on the limit of large $d$. From Eqs.~(\ref{eq:qrelsforvp0}), (\ref{eq:qd1}), (\ref{eq:qs0}) and (\ref{eq:qs1}) $v'_0$ can approximately be written as:
\begin{equation}
v'_0\approx \frac{1}{d^{3/2}}\left(1-\sqrt{\frac{2(1-x_0^2)}{d}}\frac{1}{x_0}\right)\frac{x_0x_1^2x_2v_0U^2}{x_0^2v_0^2+x_1^2U^2}.
\label{eq:v0p}
\end{equation}
Here we have neglected terms of order $d^{-1}$ times leading order and higher. Then one can write the approximate value for the violation given in Eqs.~(\ref{eq:qhalfqs1}) and (\ref{eq:qs1}) as:
\begin{equation}
Q(d)\approx \frac{1}{d}\left(1-\sqrt{\frac{8(1-x_0^2)}{d}}\frac{1}{x_0}\right)\frac{x_0^2x_1^2x_2^2v_0^2U^2}{x_0^2v_0^2+x_1^2U^2}.
\label{eq:Qdappra}
\end{equation}
If one expresses $v_0^2$ from Eq.~(\ref{eq:normapp}) and substitutes it into the equation above one gets:
\begin{equation}
Q(d)\approx \frac{1}{d^2}\left(1-\sqrt{\frac{8(1-x_0^2)}{d}}\frac{1}{x_0}\right)\frac{x_1^2x_2^2(1-2U^2d)U^2d}{1+\frac{x_1^2-x_0^2}{x_0^2}U^2d}.
\label{eq:Qdappr}
\end{equation}
From the condition that the partial derivative of the expression above in terms of $z\equiv U^2d$ is zero it follows that
\begin{equation}
z\equiv U^2d=\frac{x_0^2}{x_1^2-x_0^2}\left(\sqrt\frac{x_1^2+x_0^2}{x_0^2}-1\right).
\label{eq:U2dx}
\end{equation}
When calculating the optimum values for $x_1$, $x_2$, and $x_3$ let us neglect the term proportional to $d^{-5/2}$. If we write $x_i$ in terms of polar coordinates and demand that the partial derivatives of Eq.~(\ref{eq:Qdappra}) without the neglected term in terms of the polar angles are zero, we get, for the optimal values of $x_i^2$,
\begin{align}
x_0^2&=\frac{\sqrt{z(1-z)}-z}{2(1-2z)}\nonumber\\
x_1^2&=\frac{1-z-\sqrt{z(1-z)}}{2(1-2z)}\nonumber\\
x_2^2&=\frac{1}{2}.
\label{eq:xU2d}
\end{align}
Substituting $x_0$ and $x_1$ from the equation above into Eq.~(\ref{eq:U2dx}), after straightforward steps we can get that the optimal $z\equiv U^2d$ for large $d$ satisfies the third-order equation $4z^3-8z^2+6z-1=0$. This equation has a single real root, which is
\begin{equation}
z\equiv U^2d=\frac{1}{6}\left(4+\sqrt[3]{3\sqrt{33}-17}-\sqrt[3]{3\sqrt{33}+17}\right)\approx 0.2282.
\label{eq:Uanal}
\end{equation}
Then, from Eqs.~(\ref{eq:Qdappr}), (\ref{eq:xU2d}), and (\ref{eq:Uanal}) we can get
\begin{equation}
Q(d)\approx \frac{0.01686}{d^2}\left(1-\frac{6.118}{\sqrt{d}}\right).
\label{eq:Qdapprnum}
\end{equation}
We note that the parameters determined here are only optimal for really large $d$. For $d\leq 37$ the equation above does not even give a violation.

\end{document}